%% file: main.tex
\documentclass[journal]{IEEEtran}
\usepackage{url}
\usepackage{graphicx}
\usepackage{float}
\usepackage{xcolor}
\usepackage{braket}

\usepackage{cancel}

\ifCLASSINFOpdf
\else
\fi
\usepackage{amsmath,amssymb}

\begin{document}
%
\title{The Wigner-Ville Transform as an Information Theoretic Tool in Radio-frequency Signal Analysis}
%
%
%


\author{\IEEEauthorblockN{Erik~Lentz\IEEEauthorrefmark{1}, 
Emily~Ellwein\IEEEauthorrefmark{1}, 
Bill~Kay\IEEEauthorrefmark{1}, 
Audun~Myers\IEEEauthorrefmark{1}, 
Cameron~Mackenzie\IEEEauthorrefmark{1}} \\
\IEEEauthorblockA{\IEEEauthorrefmark{1}Pacific Northwest National Laboratory, Richland, WA 99354 USA}}
\maketitle

\input{sections/Abstract}


%
\IEEEpeerreviewmaketitle

\input{sections/Introduction}

\input{sections/Formulation}
\input{sections/Exemplars}

\input{sections/Summary}

\input{sections/Acknowledgements}

\appendices

\input{appendicies/Wigner_Alternates}

\input{appendicies/Numerical_Methods}
\input{appendicies/Autoencoder_Implementation}
\input{appendicies/Further_Results}

\ifCLASSOPTIONcaptionsoff
  \newpage
\fi



%

\newpage

\bibliographystyle{IEEEtran}
\bibliography{bibliography}

\end{document}

%% file: sections/Abstract.tex
\begin{abstract}

This paper presents novel interpretations to the field of classical signal processing of the Wigner-Ville transform as an information measurement tool. The transform's utility in detecting and localizing information-laden signals amidst noisy and cluttered backgrounds, and further providing measure of their information volumes, are detailed herein using Tsallis' entropy and information and related functionals. Example use cases in radio frequency communications are given, where Wigner-Ville-based detection measures can be seen to provide significant sensitivity advantage, for some shown contexts greater than 15~dB advantage, over energy-based measures and without extensive training routines. Such an advantage is particularly significant for applications which have limitations on observation resources including time/space integration pressures and transient and/or feeble signals, where Wigner-Ville-based methods would improve sensing effectiveness by multiple orders of magnitude. The potential for advancement of several such applications is discussed.

\end{abstract}

%% file: sections/Introduction.tex
\section{Introduction}
\label{sec:intro}

The Wigner-Ville transform (WVT) has been a deeply insightful tool in multiple fields of applied analysis for nearly a century. The transform is primarily attributed to Eugene Wigner, who constructed it as a means to describe the phase-space behavior of wave-function mechanics in the then relatively new quantum theory~\cite{Wigner1932}. Similar forms can also be found in the works of Hermann Weyl, Werner Heisenberg, and Paul Dirac of that time~\cite{Weyl_1927, Dirac_1930,Heisenberg_1931}, though these mentions missed the significance of the symplectic representation. The interpretation of the transformed function as a quasi-probability distribution according to the Born interpretation of wave-function mechanics provided a means of generalizing the classical statistical mechanics of Bernoulli-Boltzmann-Maxwell-Gibbs to pure and mixed quantum states~\cite{Moyal_1949,10.1063/1.1705323} and which include regions of negative weight. Quantum deviations from the classical phase-space distribution interpretation are due to correlations between the incompatible (inseparable) degrees of phase space, fundamentally due to the Heisenberg uncertainty principle of quantum theory. The function has become a mainstay of research in applied quantum theory, with modern uses in fields including but not limited to materials science, chemistry, sensing, computing and information sciences, and astrophysics \& cosmology~\cite{WignerSTbook2018,Weinbub2018,Lentz2019,Lentz2020}.

In the late 1940's the WVT was introduced to classical analytic signal analysis by Jean Ville~\cite{Ville_1948} 
as a means to generate a quasi-instantaneous spectrogram from a time-sampled waveform. This composite (time-frequency) transformation is an outgrowth of earlier work in composite functions by John Carson \& Thornton Fry~\cite{Carson_1937} and Dennis Gabor~\cite{Gabor_1946_231} made to address questions of harmonic analysis that were found to be insufficiently addressed by global techniques such as Fourier analysis. The WVT has been used extensively in classical harmonic analysis for analyzing non-stationary signals~\cite{Claasen1980a,Claasen1980b,Claasen1980c} with applications including in human speech~\cite{CHESTER1984,Velez1989,Hinrichs2021}, analog and digital signal processing in radio-frequency (RF) communications~\cite{Hlawatsch1992,Flandrin2002,scholl2021}, medical analysis~\cite{KITNEY1987,Chouvarda2003,Mainardi2004}, and economics~\cite{Lerner2016}, being used primarily as a generalized intensity spectrogram and then almost exclusively with major modifications. These modified forms of the WVT are designed to attenuate cross-correlations of a waveform, which are potentially present at all scales and become unavoidable at the Gabor limit, the signal analysis analogue of the Heisenberg uncertainty limit.

This paper seeks to demonstrate a broader utility and several advantages of the full WVT in classical signal analysis by taking more information-theoretic cues from the original quantum use case. The remainder of the paper is structured as follows: 
Section~\ref{sec:theory} presents the definitions of the WVT, its properties, and the information theory concepts of import; 
Section~\ref{sec:examples} demonstrates several advantageous uses of the WVT in the field of radio frequency (RF) communications amidst noisy and cluttered background environments, including enhanced detection sensitivity, localization of independent and shared information, and measurement of data volume; Section~\ref{sec:summary} summarizes the above collected insights and suggests additional classical signal analysis applications and research avenues for the WVT.

%% file: sections/Formulation.tex
\section{Formulation}
\label{sec:theory}

The (continuous) composite WVT of classical analysis is a Fourier decomposition of the $\tau$ delay-time response of a time-wise dataset's two-point auto-covariance function
\begin{equation}
    W_{\rho}(t,\nu) = \frac{1}{N_{\rho}}\int_{-\infty}^{\infty} d \tau e^{-2 \pi i \nu \tau} \rho_{(2)} (t + \tau/2, t - \tau/2), \label{eqn:WVT0}
\end{equation}
where the auto-covariance function $\rho_{(2)}$ in quantum theory is referred to as a two-point density-of-states function. The form simplifies for a single known data waveform 
\begin{equation}
    W_{\Psi}(t,\nu) = \frac{1}{N_{\psi}}\int_{-\infty}^{\infty} d \tau e^{-2 \pi i \nu \tau} \Psi^{*} (t + \tau/2) \Psi (t - \tau/2), \label{eqn:WVT}
\end{equation}
where $\Psi$ is the real- or complex-valued time-wise waveform and is taken to be of zero mean and unit size via the L$^2$ measure
\begin{equation}
    \int_{-\infty}^{\infty} dt \Psi^*(t) \Psi(t) = 1.
\end{equation}
Unless otherwise stated, the space of waveforms considered here will be continuous unit-normalized functions. This form would be referred to as a ``pure state'' transform in the quantum theory literature. The resulting function $W_{\Psi}$ has several notable properties, such as being real-valued, normalized in the presence of a normalized waveform, and under the appropriate projection may be interpreted as a time-wise or frequency-wise intensity distribution according to the Born rule
\begin{align}
    \rho_{\Psi}(t) &= \int_{-\infty}^{\infty} d \nu W_{\Psi}(t, \nu), \\
    \tilde{\rho}_{\Psi}(\nu) &= \int_{-\infty}^{\infty} d t W_{\Psi}(t, \nu).
\end{align}
Further, despite the WVT being non-linear, the Weyl transform can be used as an effective inverse to the WVT by restoring the waveform auto-correlation function via
\begin{align}
    \Phi(W_{\Psi}) (t, t') &= 2 \int_{-\infty}^{\infty} d \nu e^{2 \pi i \nu (t - t')} W_{\Psi}(t/2 + t'/2,\nu) \\
    &= \Psi^* (t) \Psi(t') , \nonumber
\end{align}
where the original waveform shape may be recovered by holding either $t$ or $t'$ constant, implying that the WVT transform is lossless to the shape of the waveform. 

The transformed function is referred to as a quasi-distribution in the quantum theory literature as it may take on negative values and therefore cannot be interpreted as a true probability distribution function (PDF). The volume measure of these negative regions in the state space of complimentary variables $t$-$\nu$ are on the order of the Heisenberg uncertainty in the quantum language or the Gabor uncertainty in signal processing. The uncertainty area is limited to be no smaller than $1/(4 \pi)$ in the given units of Eqn.~\ref{eqn:WVT}. Stated in terms of the waveform variance, the limit is
\begin{equation}
    Var_{\Psi}(\nu) \cdot Var_{\Psi}(t) \le \frac{1}{(4 \pi)^2}.
\end{equation}
The resolution limits implied by this uncertainty principle also provide a natural upper limit to the size of the symplectic state space of allowed Wigner functions and scales like the waveform's original state space~\cite{Manfredi2000,Baraniuk2001}.

Quantifying the (in)dependence of waveforms via their WVT quasi-distributional occupation of the $t$-$\nu$ symplectic state space requires some elaboration. Waveforms in the use cases considered here are composed of additively superposed components, with independence or correlation of constituents determined by local overlap using the L$^2$-inspired inner product
\begin{equation}
    \braket{x,y} =  \int_{-\infty}^{\infty} d \tau x^* (\tau) y (\tau).
\end{equation}
Therefore, given the additive superposition of two arbitrary unnormalized waveforms $\Theta$ and $\Phi$ of the form $X = \Theta + \Phi$, their also unnormalized WVT decomposes to
\begin{align}
    &W_{X}(t,\nu) = W_{\Theta}(t,\nu) + W_{\Phi}(t,\nu) \nonumber \\
    &+ 2 \mathrm{Re} \int_{-\infty}^{\infty} d \tau e^{-2 \pi i \nu \tau} \Theta^{*} (t + \tau/2) \Phi (t - \tau/2), \label{eqn:WVTtwosignals} 
\end{align}
where pure additivity in the WVT occurs if the cross terms between $\Theta$ and $\Phi$ vanish, which also corresponds to $\braket{\Theta,\Phi} = 0$. Under these conditions, the WVTs of the individual components are also found to be independent as measured by $\int dt d\nu W_{\Phi}(t,\nu) W_{\Theta}(t,\nu)$.

\begin{figure}[h!]
\begin{center}
    \includegraphics[width=1.0\columnwidth]{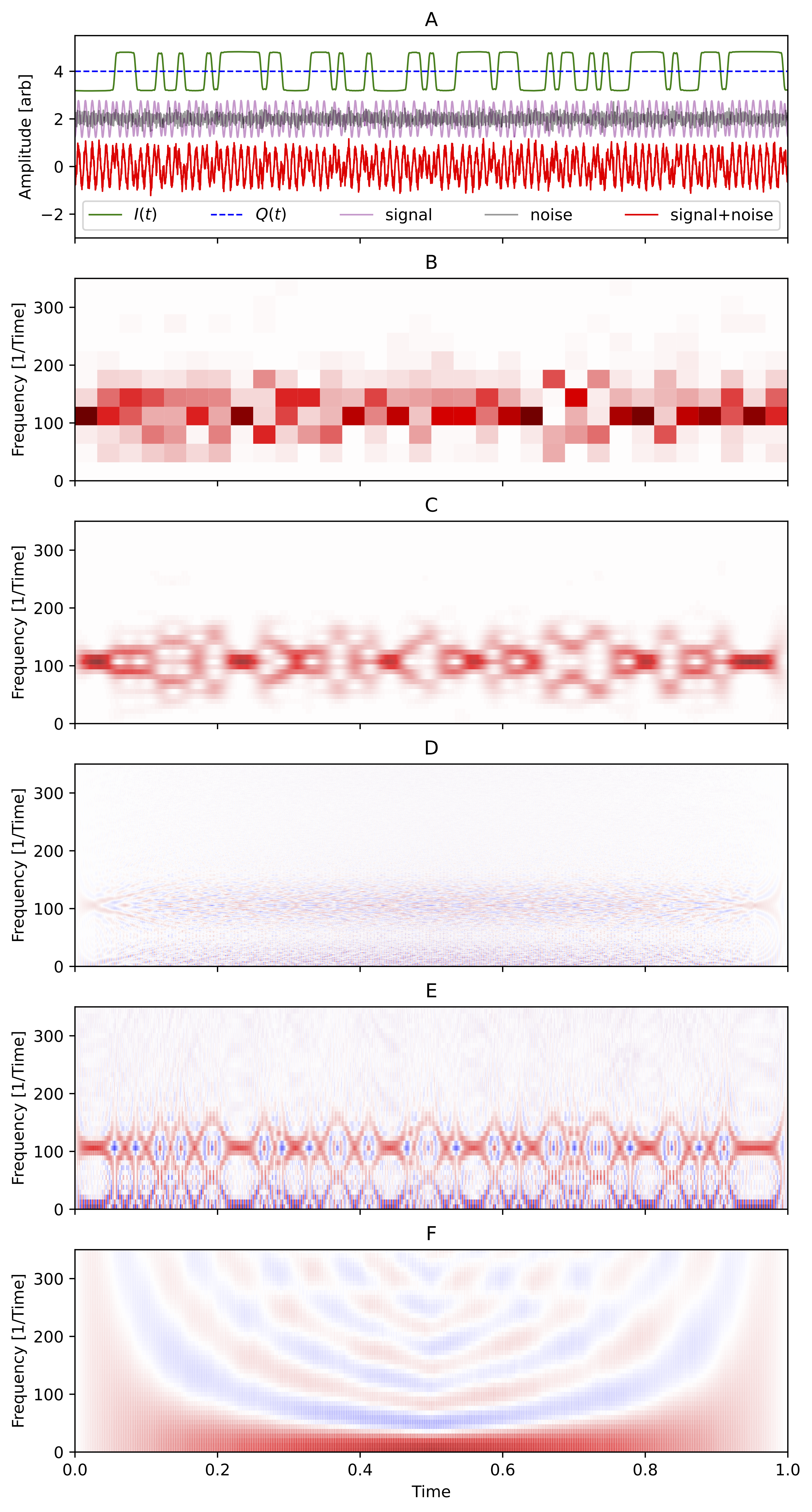}
    \caption{An example RF signal and sample resulting Wigner-Ville-type Cohen's class transformations. Only the non-negative-frequency half-plane is shown, though each of the transforms is symmetric about zero. The signal is composed of an additive superposition of white Gaussian noise and a BPSK transmission with root-raised-cosine filter with shape parameter RRC=0.5, containing a message of random symbols, injected with a (log) signal-to-noise ratio (SNR) of 10. Alternatives to the WVT are referenced in Appendix~\ref{appx:wigner_alternates}. The length of the waveform is $N = 2^{12} = 4,096$ samples, $N_s = 100$ is the number of symbols.
    (A) Time domain representation of the waveform components, vertically offset, with the I-Q representation of the BPSK message at top, individual noise and transmission components ad middle, and superposed waveform at bottom.
    (B) The spectrogram of the waveform with FFT windows of size $N_{FFT} = 2^7 = 128$. 
    (C) The modulus squared of the Gabor time-frequency representation using smoothing length of $L_s = 60$. 
    (D) The full WVT of the waveform. 
    (E) The pseudo-Wigner transform with Gaussian smoothing kernel of width $\sigma = 60$,
    (F) Polynomial Wigner with smoothing length $\sigma = 60$, $q = 4$, $b = [2,0,0,0,-2]$, and $c = [1,0,0,0,-1]$. All intensities are plotted in linear units.
    }
    \label{fig:signal}
\end{center}
\end{figure}

As has already been hinted at, the cross-correlation contributions to a WVT do not necessarily correspond to regions of energy intensity as in a traditional spectrogram, and therefore have most often been considered as an undesirable side-effect of the transform. Several modified versions of the WVT have been presented in the literature primarily to better resemble the power spectrogram 
including  Gabor-Wigner transforms~\cite{Pei2007}, 
polynomial Wigner transforms~\cite{Boashash1992,Boashash1994}, 
and windowed Wigner transforms such as from the Cohen's class~\cite{Cohen1995},
definitions of which are provided in Appendix~\ref{appx:wigner_alternates} for reference. These alternate transforms are not invertable in general, smoothing structures essential to the quantum use case and blunting the transform's efficacy. An illustration of the bare and modified WVTs on an example noisy Binary Phase-shift Keying (BPSK) modulated transmission is provided in Fig.~\ref{fig:signal}. 

This work seeks to exploit the full structure of the WVT-ed waveforms for classical signal analysis. The tools put forth in the remainder of this section concentrate on answering questions of information content in waveform data and are inspired by both the classical and quantum information theory literature. 

Information is taken to be the resolution of uncertainty leading to interpretation and meaning, and information can be imprinted to a waveform in numerous ways within the span of that waveform's state space. Even with the WVT's compression-less quasi-distribution representation of that waveform in the $t$-$\nu$ plane, the meaning of a waveform may still prove evasive without the key of modulation. Instead of pursuing the specifics of the encoded information, this work focuses on revealing quantities independent of modulation such as information volume \& localization, and mutual information. Entropy is often considered as a complimentary quantity to information, however the usual real-logarithmic form of Shannon is incompatible with the negative values of the WVT quasi-distribution. Instead, this work takes inspiration from the Tsallis class of entropy functionals~\cite{Tsallis1988}, which when applied to a countable set of state probabilities $\{ p_i \}_{i \in A}$ takes the form
\begin{equation}
    S_{\alpha} = \frac{1 - \sum_i p_i^{\alpha}}{\alpha-1}, 
\end{equation}
where the Shannon form is recovered in the limit $\alpha \to 1$. This work in particular will concentrate on order $\alpha = 2$ Tsallis entropy (TE) which fortuitously has the compact integral form when using the WVT
\begin{equation}
    S_{2}(W_{\Psi}) = 1 - \int dt d\nu W_{\Psi}^{2}(t,\nu),
\end{equation}
where the Tsallis information (TI) content is given as the compliment, reminiscent of the L$^2$ measure over the $t$-$\nu$ plane
\begin{equation}
    I_2(W_{\Psi}) = 1 - S_2(\Psi) = \int dt d\nu W_{\Psi}^{2}(t,\nu).
\end{equation}
Interpretation of the WVT-TE/TI compositions as entropy/information measures is not as clear as with Shannon (number of bits needed to store a state in the given representation) and requires additional exposition on their properties. 

\begin{figure}[h!]
\begin{center}
    \includegraphics[width=\columnwidth]{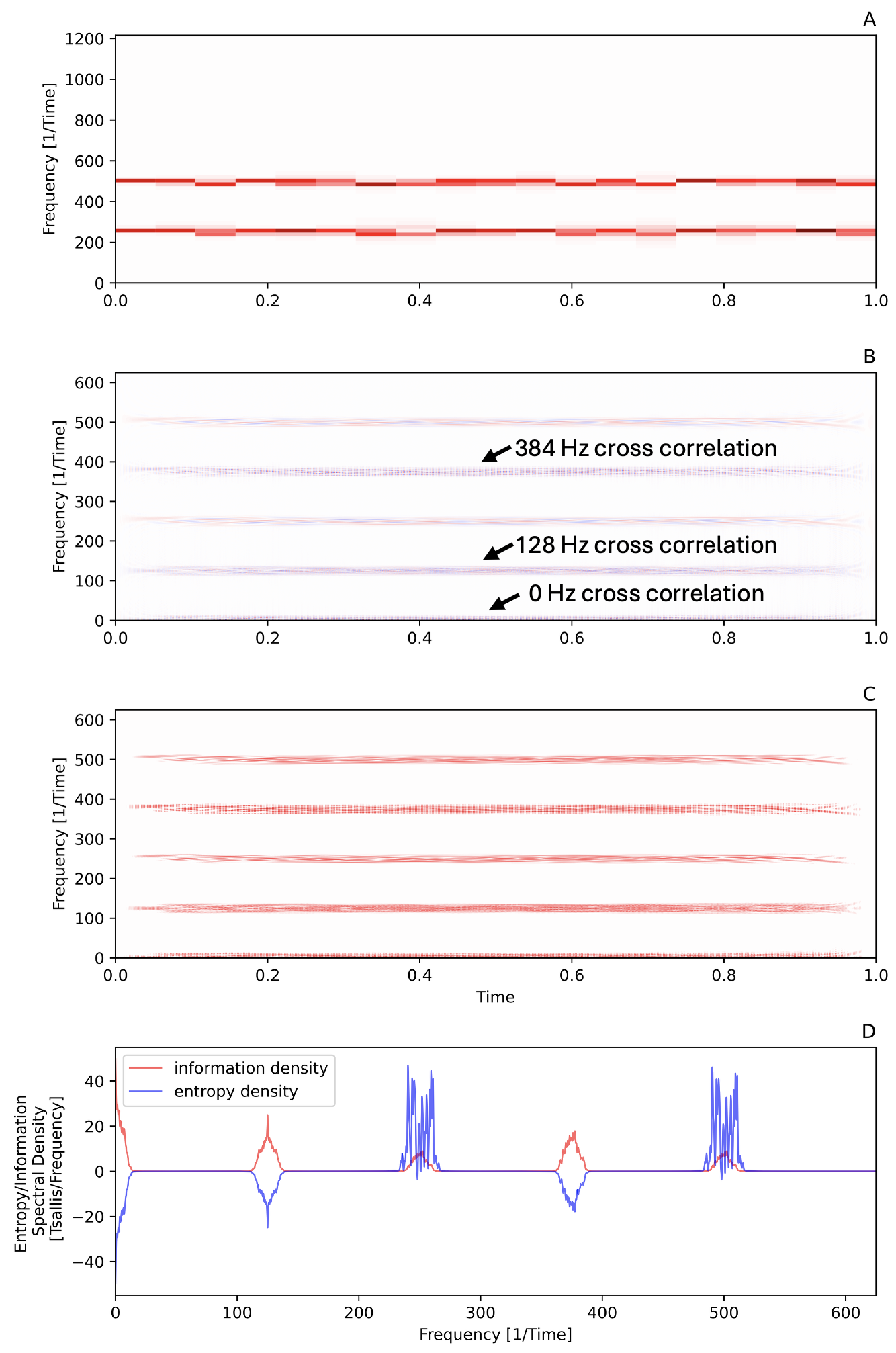}
    \caption{Example properties of WVT, information and entropy densities on a noise-less waveform containing two QPSK signals with identical messages (RRC = 0.35, $R_{\text{Symb}} = 100$~Symb/s) collected at a rate of 10,000. 
    (A) The waveform spectrogram with $N_{FFT} = 128$. 
    (B) The full WVT of the waveform. 
    (C) The information density $i_2(t,\nu)$ of the waveform. 
    (D) The projected information density $I_{2,\nu}$ (red) and entropy density $S_{2,\nu}$ (Blue) of the waveform. All intensities are plotted in linear units.
    }
    \label{fig:statprop}
\end{center}
\end{figure}

The WVT given in Eqn.~\ref{eqn:WVT} is known as an idempotent representation of a unit-normalized input waveform, meaning that the trace over the WVT-squared also evaluates to unity
\begin{equation}
    \int dt d\nu W_{\Psi}^2(t,\nu) = 1,
\end{equation}
regardless of the normalized waveform's content. This implies that the total entropy and information evaluate to $S_2 = 0$ and $I_2 = 1$ respectively, reflecting complete resolution of uncertainty and meaning of the waveform state. The localized measures of entropy and information will be far more enlightening tools
\begin{align}
    s_2(W_{\Psi})(t,\nu) &= W_{\Psi}(t,\nu) - W_{\Psi}^2(t,\nu), \\
    i_2(W_{\Psi})(t,\nu) &= W_{\Psi}^2(t,\nu),
\end{align}
with the densities over time or frequency subspaces being
\begin{align}
    S_{2,t}(W_{\Psi}) &= \int d\nu W_{\Psi}(t,\nu) - \int d\nu W_{\Psi}^2(t,\nu), \\
    S_{2,\nu}(W_{\Psi}) &= \int dt W_{\Psi}(t,\nu) - \int dt W_{\Psi}^2(t,\nu), \\
    I_{2,t}(W_{\Psi}) &= \int d\nu W_{\Psi}^2(t,\nu), \\
    I_{2,\nu}(W_{\Psi}) &= \int dt W_{\Psi}^2(t,\nu),
\end{align} 
each of which will be interpreted as a relative share of total TE/TI carried by the waveform. The TE/TI forms contain multiple additional properties desirable for entropy/information forms which will be used throughout this work, with Fig.~\ref{fig:statprop} displaying several of these behaviors. 
\begin{itemize}
    \item \textbf{Concavity/Convexity in distributions:} The entropy functional $S$ of (quasi-)distributions $d_1$, $d_2$ in convex combination will be concave $S(\lambda d_1 + (1- \lambda) d_2) \ge \lambda S(d_1) + (1- \lambda) S(d_2)$. The information functional $I$ exhibits convex behavior $I(\lambda d_1 + (1- \lambda) d_2) \le \lambda I(d_1) + (1- \lambda) I(d_2)$. The entropy/information densities also contain the respective convex/concave behavior.
    \item \textbf{Additivity for independent random variables:} The global entropy functional of independent random variables, given by waveforms  $A$ and $B$, is additive $S_2(W_{A+B}) = S_2(W_{A}) + S_2(W_{B})$. The global information functional $I_2$ is also additive. The entropy/information densities require stricter conditions on independence, which will be introduced as needed.
    \item \textbf{Sub-additivity for dependent variables:} The entropy functional of dependent random variables is less than the sum of the individual subsystems \\
    $S_2(W_{A+B}) \le S_2(W_{A}) + S_2(W_{B})$ with additivity only being restored when $A$ and $B$ become independent. The information functional conversely is super-additive. The entropy/information densities again require stricter conditions, which will be introduced as needed.
    \item \textbf{Expansibility:} The entropy and information functionals are unchanged by adding outcomes with null likelihood.
    \item \textbf{Symmetry:} The entropy and information functionals are invariant under argument permutation of input distributions.
\end{itemize}

Cross-correlation effects are able to capture crucial features of shared information between components. The quadratic order in the waveform allows for different time and frequency components to interfere, providing a measure of the cross-time and/or cross-frequency correlations localized at the components' midpoint. Nominally, these interference effects are straightforward to locate, but as will be seen in the next section in the presence of crowded waveforms composed of multiple sources, the power spectra and interference terms can overlap and become more difficult to isolate. Figure~\ref{fig:statprop} demonstrates some typical behaviors of the Tsallis information/entropy measures on a waveform of duration one second and sampled at a rate of 2.5~kSamp/s that contains two Quadrature Phase-shift Keying (QPSK) signals at carriers $f_1 = 256$~Hz and $f_2 = 512$~Hz, both with information transmitted via evolving phase as $\phi(t)$ with symbol rate of $R_{\text{Symb}} = 25$~Symb/s (100~Samp/Symb). As can be gleaned from Eqn.~\ref{eqn:WVTtwosignals}, the WVT of this waveform presents power-spectral-density (PSD)-like lines centered about $\pm f_1$ and $\pm f_2$ (though only the positive-frequency components are shown in the figure) given by the WVT of the individual QPSK transmissions, as well as cross terms that in the limit of slow phase evolution is seen to be most prominent about the midpoint frequencies of each signal combination 
$\bar{f}_{\texttt{cross}} = 
\pm (f_1 - f_1)/2, 
\pm (f_2 - f_2)/2, 
\pm (f_1 + f_2)/2, 
\pm (f_1 - f_2)/2 $ 
with phase that evolves respectively as 
$\phi_{\texttt{cross}} = 
\pm 2 \pi t (2 f_1) \pm 2 \phi_1, 
\pm 2 \pi t (2 f_2) \pm 2 \phi_2, 
\pm 2 \pi t (f_1 - f_2) \pm (\phi_1 - \phi_2), 
\pm 2 \pi t (f_1 + f_2) \pm (\phi_1 + \phi_2)$. 
Laid out more plainly for the signals in Fig.~\ref{fig:statprop}, the cross terms are centered at frequencies $\bar{f}_{\texttt{cross}} = \pm 0, \pm 128, \pm 384$~Hz. The entropy/information densities capture the regions of high power and also the shared information at their median. The zero-frequency line may be used as a means of gauging total captured information due to the comparison of positive frequency components with their mirrored negative frequency components. The volume of information shows the individual QPSK transmissions carry equal amounts of information, but when integrated with the interference term, reveals the presence of shared information. 

%% file: sections/Exemplars.tex
\section{Use Case: Radio-frequency Communications Sensing}
\label{sec:examples}

This section highlights several motivating capabilities of a WVT-based signal analysis applied to the passive sensing of radio-frequency (RF) communications signals where the analysis has limited prior knowledge of time-resolved waveform contents. The target capabilities are the accurate detection, localization, and information volume estimation of injected signals. Performance of the WVT-based analyses are compared to standard tools of the field including spectrogram-based detection and established machine learning techniques where applicable. Numerical implementations of the WVT methods are provided in Appendix~\ref{appx:discrete}.

The technology of transmitting information via a time-wise waveform using an agreed-upon interpretation protocol between transmitter and receiver has been central to (tele-)communications for centuries. 
Telecommunications is still based on this paradigm, though its components have become much more sophisticated, allowing for high-speed fideletous transmission and reception over long distances and through multiple mediums. The count and complexity of waveform modulation protocols have grown enormously, particularly in the digital space, to suit various objectives such as increased transmission speed, security, etc.. The advent of AI-driven protocol to tackle optimization of bandwidth, usage, and other objectives may see this trend accelerate.

The first demonstration considers signals composed of a superposition of additive white Gaussian noise (AWGN) background and an isolated modulated signal. The modulations to be shown in this example are On-off Keying (OOK), BPSK, 64-Quadrature Amplitude Modulation (64QAM), and Multi-frequency Shift Keying (MFSK). Each modulation has a carrier frequency of $f_c = 1,000$~Hz save for MFSK that randomly switches between carrier frequencies every 50 symbols, and each modulation uses a root-raised-cosine filter with  RRC=0.35. The waveforms are sampled at a rate of 10~kSamp/sec over the unit integration time, making each waveform 10kSamp long. The symbol rate parameter is explored at values of $R_{\text{Symb}} \in \{ 1,5,25,100,250 \}$~Symb/sec (10k,2k,400,100,40~Samp/Symb, respectively) for all modulations. Integration time is one second. These modulations and parameters were chosen to sample a range of keying approaches, constellation sizes, and spectral widths. 

Classifying the presence or absence of an injected signal from the waveform's WVT or other derived measures from Sec.~\ref{sec:theory} are performed here using simple threshold techniques explained below. This approach is chosen for ease of interpretation and explainability. Combining the Wigner-based measures with more complex classifiers is deferred to future work. 

\begin{itemize}
    \item \textbf{$I_{2,\nu}$ entropy:} The TI spectral density of a normalized waveform is interpreted as a distribution, then used as input to compute the Shannon entropy $H = \sum p_i \ln{ p_i }$. The threshold for detection is given by the $p_{FA}$ lower quantile boundary from a population of $S(I_{2,\nu} d \nu )$ computed from a collection of background-only baseline measurements.  

    \item \textbf{$S_{2,\nu}$ entropy:} The positive-value regions of the TE spectral density of a normalized waveform is interpreted as a distribution, then used as input to compute the Shannon entropy. The threshold for detection is given by the $p_{FA}$ upper quantile boundary from a population of $H(S_{2,\nu} d \nu )$ computed from a collection of background-only baseline measurements.

    \item \textbf{$\rho_{\nu}$ `entropy':} The WVT spectral density of a waveform is used as input to compute the Shannon entropy. The threshold for detection is given by the $p_{FA}$ upper quantile boundary from a population of $H(\rho_{\nu} d \nu )$ computed from a collection of normalized background-only baseline measurements. Note that the Shannon measure is no longer bounded due to the relaxation of waveform normalization.

    \item \textbf{$WVT^2_{\nu}$ `entropy':} The TI spectral density of a waveform is used as input to compute the Shannon entropy. The threshold for detection is given by the $p_{FA}$ upper quantile boundary from a population of $H(WVT^2_{\nu} d \nu )$ computed from a collection of normalized background-only baseline measurements. Likewise, the Shannon measure here is no longer bounded.

\end{itemize}

Several other classifiers based on the waveform PSD are also provided to serve as benchmarks.

\begin{itemize}
    \item \textbf{PSD peak:} The spectral distribution's largest positive peak is identified as a signal candidate. The threshold for detection is given by the $p_{FA}$ upper quantile boundary from a population of PSD maximum peaks computed from a collection of normalized background-only baseline measurements.

    \item \textbf{PSD entropy:} The PSD of an normalized waveform is interpreted as a distribution, then used as input to compute the Shannon entropy. The threshold for detection is given by the $p_{FA}$ upper quantile boundary from a population of PSD entropies $H( |\tilde{\Psi}|^2 d \nu)$ computed from a collection of normalized background-only baseline measurements.

    \item \textbf{AE-TS:} Autoencoder (AE)-based neural networks are trained on the time series from a collection of background-only baseline measurements. Anomalies identified through significantly higher reconstruction errors indicating deviation from learned baseline patterns are used as the means of injected signal detection. Implementation details for the AE used are provided in Appendix~\ref{appx:ae}.

    \item \textbf{AE-PSD:} Autoencoder (AE)-based neural networks are trained on the PSDs from a collection of background-only baseline measurements. Anomalies identified through significantly higher reconstruction errors indicating deviation from learned baseline patterns are used as the means of injected signal detection. Implementation details for the AE used are provided in Appendix~\ref{appx:ae}. 
    
\end{itemize}

The AWGN-only background in this first demonstration provides a Johnson-Nyquist-like frequency spectrum that is nearly homogeneous with perturbations for both the PSD and WVT representations. A uniform distribution produces maximum Shannon entropy, with any shape change in the distribution suppressing that value, hence its choice for use in this demonstration.

\begin{figure}[ht!]
\begin{center}
    \includegraphics[width=1.0\columnwidth]{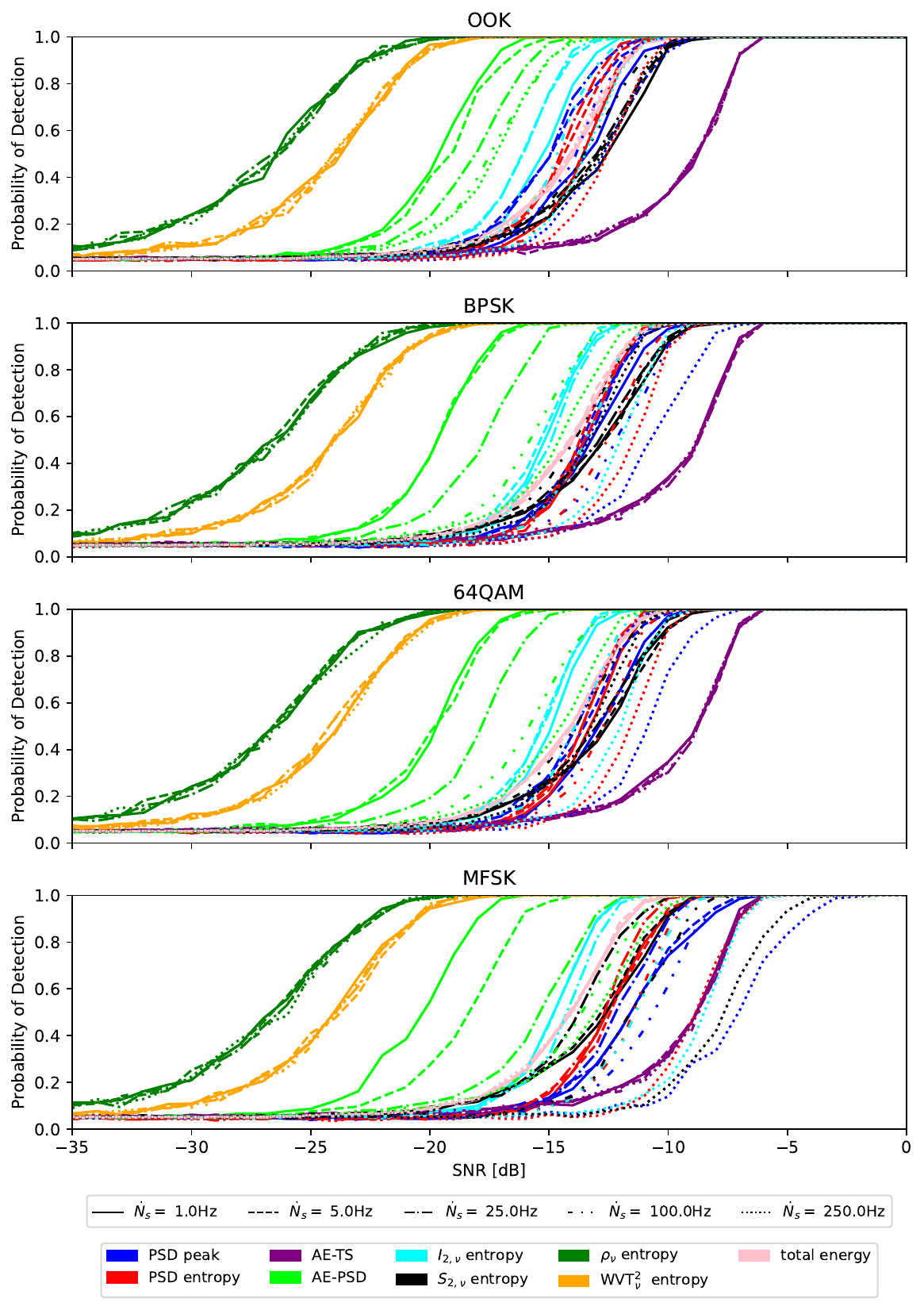}
    \caption{Global detection rate for common modulation types injected over AWGN. The detection threshold in each metric was computed in the no-injection limit using a probability of false alarm $p_{FA} = 0.05$, and was reached using $N_{\text{trials}} = p_{FA}^{-2} = 400$. The detection statistics at each point along SNR are also generated by $N_{\text{trials}}$. The pseudo-WVT-based measures use a width parameter of $\sigma=100$.}
    \label{fig:single_injection_detection}
\end{center}
\end{figure}

Probability of detection of injected signals versus SNR of WVT-based and benchmark measures using the above thresholds are shown in Fig.~\ref{fig:single_injection_detection} using a false positive likelihood of $p_{FA} = 0.05$. Detection probabilities for the energy-based (PSD) analytic measures are seen to roughly correspond to the integrated probability for a normal distribution $P_N ( Z_{p_{FA}} \le z \le Z_{p_{FA}} + SNR \sqrt{N_{obs} b_{tot}/b_{sig}})$ where $N_{obs}$ is the number of observed spectra, $b_{tot}$ is the total bandwidth of the observation spectra, and $b_{sig}$ is the signal bandwidth. Recall that the symbol rates as a fraction of carrier frequency are $R_{\text{Symb}}/f_c \in \{ 0.001,0.005,0.025,0.1,0.25 \}$ (save for MFSK which has no set carrier frequency) corresponding roughly to the signal's fractional bandwidth $b_{sig}/f_c$. The trained AE networks showed lesser sensitivity (AE-TS) or a multi-fold increase in sensitivity (AE-PSD) compared to the more analytical peak-finder and entropy methods. The advantage of the trained AE-PSD networks may be understood as the identification of multiple latent features of the signal compounded to form a more sensitive anomaly detector than by measuring injected power alone. Lastly, note the degradation of performance of all the benchmark methods between signal bandwidth and modulation complexity, particularly in the case of MFSK where the carrier frequency itself moves randomly.

The TI and TE densities $I_{2,\nu}$ and $S_{2,\nu}$ are seen to be slightly more or slightly less sensitive relative to the energy-based analytic benchmark measures, although their degradation with signals of increasing bandwidth and complexity is often less pronounced. As was noted in the Fig.~\ref{fig:single_injection_detection} caption, the pseudo-WVT of width $\sigma=100$~samples was substituted for the WVT transform. The window width does have an effect on the detection of such simple signals, though the impact becomes marginal for widths much larger than the inverse symbol rate as the injected signal's message symbols are randomly selected. It is expected that the advantage to using the full WVT will become larger for signals and backgrounds where wider correlations are present. 

Relaxing the normalization condition shows significant ($>$10~dB) sensitivity enhancement for the WVT frequency spectrum $\tilde{\rho}_{\nu}$, and WVT-squared frequency spectrum $WVT^{2}_{\nu} = \int dt W_{\Psi}^2(t,\nu)$ components of TE and TI over the analytic energy-based counterparts, and a smaller though still a notable enhancement over the AE-PSD networks. The normalization relaxation effectively reintroduces the energy-scaling feature into the time-frequency representation. This combination of both power and information appears to ultimately lead to drastically more sensitive detection than either measure alone, although the Shannon measure can no longer be strictly interpreted as an entropy. Further, both the $\tilde{\rho}_{\nu}$ and $WVT^{2}_{\nu}$ measures show negligible sensitivity to injected signal bandwidth and modulation complexity in the presence of AWGN. 

\begin{figure}[ht!]
\begin{center}
    \includegraphics[width=0.95\columnwidth]{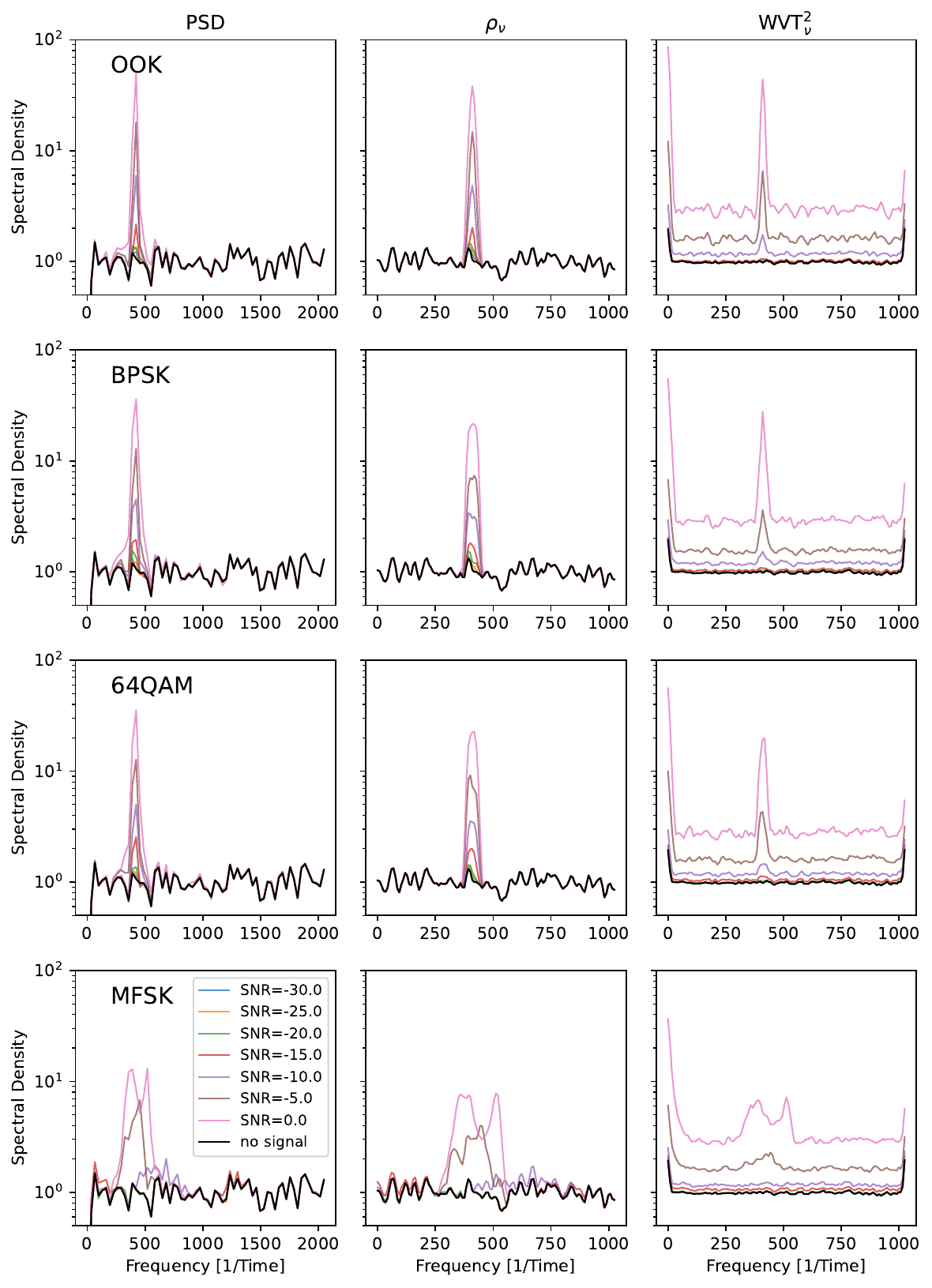}
    \caption{Localized PSD, WVT frequency spectrum $\tilde{\rho}_{\nu}$, and WVT-squared frequency spectrum $WVT^{2}_{\nu}$, the last two being the building blocks of TE and TI. Each measure's spectra is normalized to the AWGN background for better visualization of the emerging signatures. The spectra are computed using the same AWGN background to highlight the changes induced by injections of increasing intensity. The injections use a symbol rate of $R_{\text{Symb}} = 100$~Symb/s.}
    \label{fig:single_injection_localization}
\end{center}
\end{figure}

The localization of injections in the frequency domain measures of PSD, $\tilde{\rho}_{\nu}$, and $WVT^{2}_{\nu}$ for example injections with symbol rate $R_{\text{Symb}} = 100$~Symb/s and SNRs between -35~dB and 0~dB can be seen in Fig.~\ref{fig:single_injection_localization}. Each measure's spectra is normalized to the AWGN background for better visualization of the emerging signatures. The injection excesses in the PSD are well localized to a single-peaked excess for the single-carrier waveforms and scale as SNR, while the MFSK is seen to be broadly distributed leading to a more difficult detection at a given SNR as was seen in Fig.~\ref{fig:single_injection_detection}.

The WVT-based measures also create localized peaks in the frequency spectrum, similar to what was seen in Fig.~\ref{fig:statprop}. These local prominences scale with SNR in the case of $\tilde{\rho}_{\nu}$ as the WVT is of the same order in $\Psi$ as the PSD, and with SNR$^2$ in the case of $WVT^{2}_{\nu}$. The auto-covariance of the WVT allows it to further average down the variance in the AWGN while simultaneously producing a higher spectral resolution and equally prominent representation of the injected signal compared to the PSD, even when using the pseudo-WVT with a limited window width ($\sigma=60$~samples used here). This is seen in both of the shown WVT-based measures, with $WVT^{2}_{\nu}$ having the additional features of producing local prominences at $f=0$ as interference between all positive and negative frequency components of the injected signal and AWGN are combined, squared, and integrated across time, as well as a secondary prominence at the top end of the frequency spectrum. Further, the $WVT^{2}_{\nu}$ also reveals interference of the injected signal with the AWGN as a subdominant strength signature represented at all remaining frequencies, raising the overall spectra. Localization over bot time and frequency domains would be required to better localize transient signals or center-frequency-shifting signals like MFSK.

\begin{figure}[ht!]
\begin{center}
    \includegraphics[width=1.0\columnwidth]{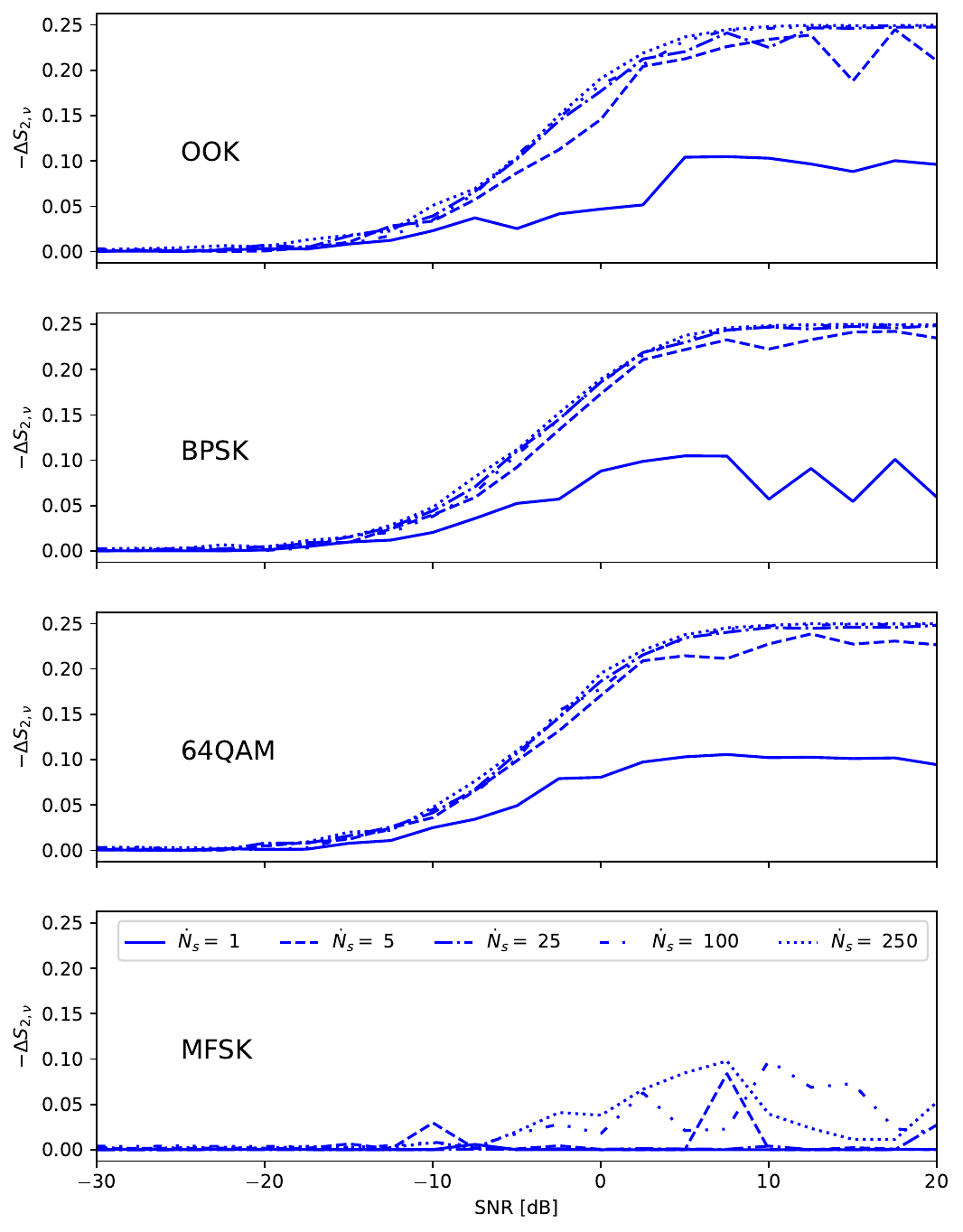}
    \caption{Relative entropy volume for example single-injection signals over AWGN. The relative measurements are made between the carrier mode and a nearby region of the same width. MFSK injections figures are largely unsettled due to lack of a well-defined prescribed center frequency. Note the significantly higher SNR required for convergence of the TE/TI volume.}
    \label{fig:single_injection_volume}
\end{center}
\end{figure}

Integrating the information or entropy spectrum can be used to estimate the total data volume transmitted within that region. In this example, integrated entropies will be compared in the vicinity of the signal $S^{\text{sig}}_{2,\nu}$ and in a background region $S^{\text{back}}_{2,\nu}$ to form a preliminary estimate of the data transmitted by the injected signal. The integrated regions are chosen around the center frequency $f_c = 1,000$~Hz with width of the symbol rate for $S^{\text{sig}}_{2,\nu}$ and a region of the same size placed 75\% up the frequency range for $S^{\text{back}}_{2,\nu}$, intended to be well away from features of the injected signal. Figure~\ref{fig:single_injection_volume} shows the entropy volume difference $- \Delta S_{2,\nu} = S^{\text{back}}_{2,\nu} - S^{\text{sig}}_{2,\nu}$. Note that this technique is sufficient for modulations with fixed carrier frequency, but poorly suited for the MFSK example injections with many carriers. This is intended to highlight the sensitivity of the information measure to location. More sophisticated integration routines that incorporate localized tracking of a signal's information are held for future work. The full WVTs were used here to represent the injected signal with full fidelity. Volume estimates required several decades higher SNR to converge compared to reliable global detection for single-carrier injected signals, while the MFSK signal did not converge as its data is poorly localized in frequency, though an accompanying time-frequency resolved localization model would significantly improve this result. Further, while this preliminary estimate of data volume is well ordered with respect to total transmitted data, the data units of Tsallis are different from Shannons. An additional mapping is required to identify the number of Shannons or bits transmitted.

A second example considers more realistic RF environments cluttered with RF-interference (RFI). Here the background of AWGN is joined by an MFSK injection with the same overall power and with symbol rate $R_{\text{Symb}} = 100$~Symb/s. The experiments of the first tests are repeated for this new background with updated thresholds. Detection sensitivity is given in Fig.~\ref{fig:multi_background_detection}. Frequency spectra and data volume estimates can be found in Appendix~\ref{appx:further}. Global detection sensitivity is seen to be compromised by the RFI beyond the overall increased power level of the background across all methods to various degrees. PSD-peak is seen to have also become highly symbol-rate-sensitive, as is PSD-entropy though to a lesser degree. The AE-TS and AE-PSD networks are similarly impacted, with the latter holding its ranking far better than the former. Total energy sensitivity is now nearly at the level of the AE-PSD networks. TI is seen to have detection rates slightly higher than the PSD-entropy level of sensitivity, while TE is overall performing far worse with a handful of positive exceptions at high symbol rates. The $\tilde{\rho}_{\nu}$ and $WVT^2_{\nu}$ methods have maintained much of their symbol-rate robustness and a significant sensitivity advantage over the other techniques, though the $\tilde{\rho}_{\nu}$ appears to have been impacted much more severely. 

\begin{figure}[ht!]
\begin{center}
    \includegraphics[width=1.0\columnwidth]{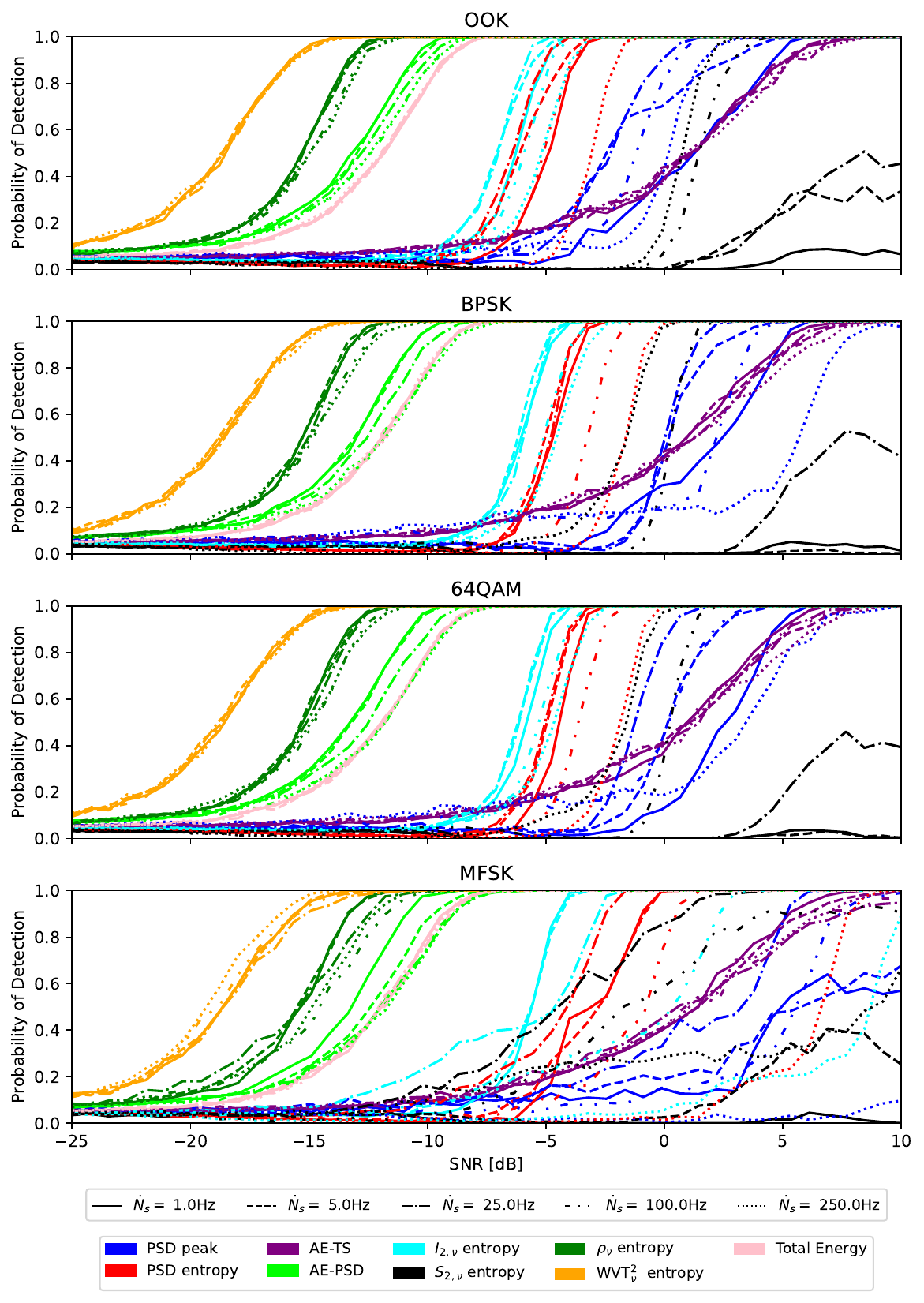}
    \caption{Global detection rate for common modulation types injected over AWGN+MFSK background. The detection threshold in each metric was computed in the no-injection limit using a probability of false alarm $p_{FA} = 0.05$, and was reached using $N_{\text{trials}} = p_{FA}^{-2} = 400$. The detection statistics at each point along SNR are also generated by $N_{\text{trials}}$. Detection rates overall were shifted to higher SNR, with higher sensitivity to symbol rate increase for all but the $\tilde{\rho}_{\nu}$ and WVT-squared frequency spectrum measures.}
    \label{fig:multi_background_detection}
\end{center}
\end{figure}

\begin{figure}[ht!]
\begin{center}
    \includegraphics[width=1.0\columnwidth]{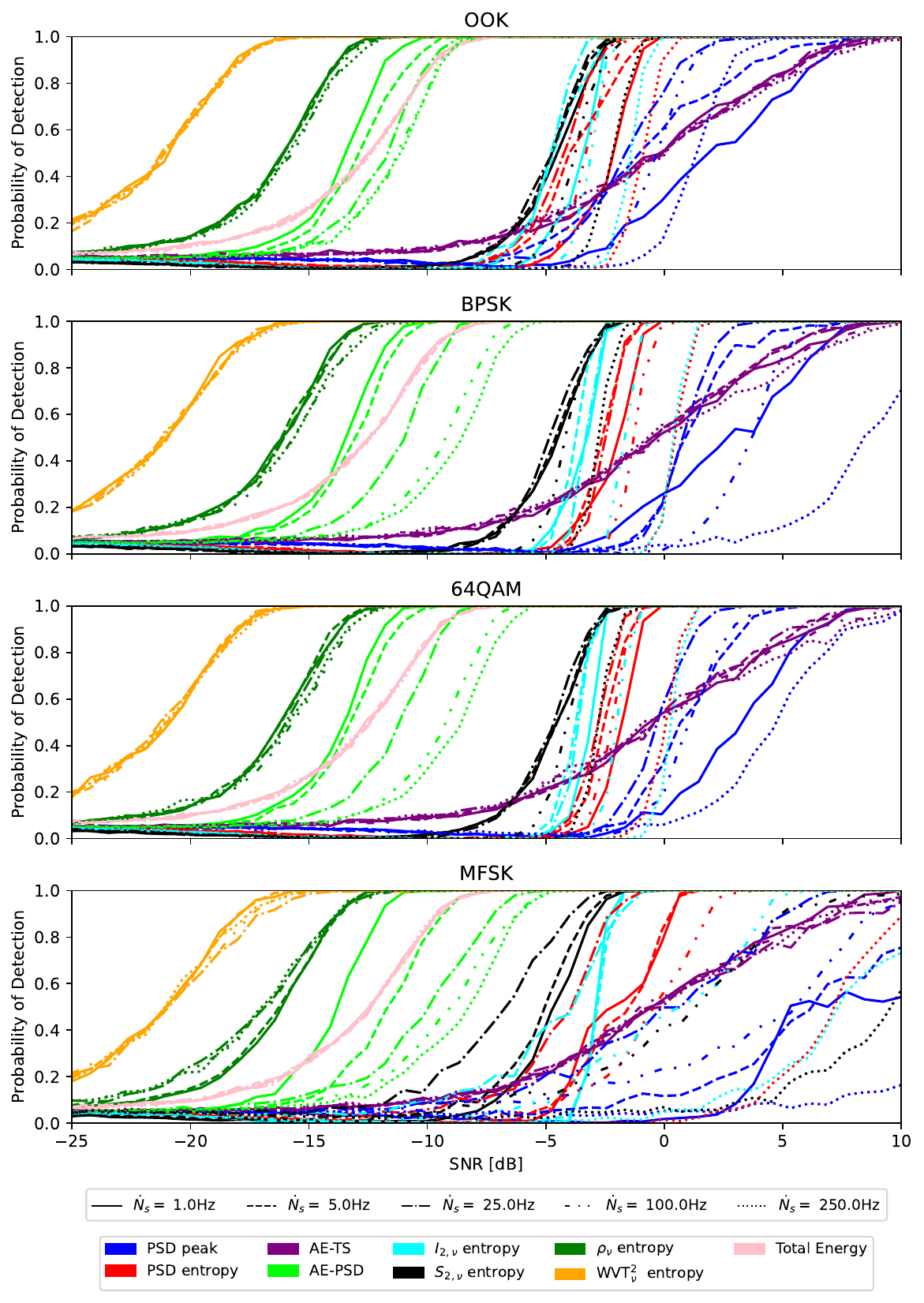}
    \caption{Global detection rate for matched injections for common modulation types injected over AWGN+(2)MFSK background. The detection threshold in each metric was computed in the no-injection limit using a probability of false alarm $p_{FA} = 0.05$, and was reached using $N_{\text{trials}} = p_{FA}^{-2} = 400$. The detection statistics at each point along SNR are also generated by $N_{\text{trials}}$. Detection rates overall were shifted to higher SNR, with higher sensitivity to symbol rate increase for all but the $\tilde{\rho}_{\nu}$ and WVT-squared frequency spectrum measures.}
    \label{fig:multi_background_detection_multi_injection}
\end{center}
\end{figure}

The third and last example discussed here further clutters the background by adding a second MFSK signal to the background (AWGN+(2)MFSK) with independent random frequencies centered around $f=2,000$~Hz, as well as a second injected signal also centered at $f=2,000$~Hz with modulation parameters and message symbols paired to the first injection. The global sensitivity of finding the matched injected signal pair amidst the AWGN+(2)MFSK background is shown in Fig.~\ref{fig:multi_background_detection_multi_injection}. Frequency spectra and data volume estimates can be found in Appendix~\ref{appx:further}. The analytic PSD methods show further degradation due to the increasingly cluttered background. TI shows some degradation on pace with the PSD-entropy method. TE shows a notable improvement in sensitivity bringing it up to or above the level of TI. The AE networks have largely maintained their sensitivity marks at low symbol rates and show more degradation at higher symbol rate than the previous AWGN+MFSK example. The $\tilde{\rho}_{\nu}$ and $WVT^2_{\nu}$ methods have become slightly more sensitive than in the less cluttered AWGN+MFSK, increasing their advantage in the more cluttered environment.

%% file: sections/Summary.tex
\section{Summary}
\label{sec:summary}

This paper has presented the features of the full Wigner-Ville transform as an information theoretic tool for classical signal analysis. Interpreting the fidelitous transform's time-frequency representation output as a quasi-distribution of a time-wise waveform's state, we leveraged the tools of information theory to extract quantities such as entropy of the Tsallis type, coherence, and independent versus shared information. These measures were applied to example use cases in the detection and characterization of radio-frequency communications signals. The WVT-based tools as global signal detectors in noisy and crowded environments showed a robustness to RF modulation type, becoming greatly more sensitive (\textgreater10~dB) than standard energy-based methods, particularly for spectrally broad signals, such as is found with fast transients and spread spectrum modulations. As a local detector, the WVT-tools are able to localize signal information similarly to energy-based techniques, and more broadly measure mutual information distributed across time and frequency domains. This information measure can be used to estimate data volume within a given domain of time-frequency, a modulation-agnostic feature not seen elsewhere in the literature without significant domain-specific training. No extensive training was required prior using the Wigner-Ville-based tools in the above use cases. 

We find these results highly motivating to explore the use of information-sensitive WVT methods to augment classical signal analysis in the numerous other fields that regularly use time/space resolved data, not only in RF communications. Further, though the tools don't require training themselves, their noise-robustness may make them highly useful in building more robust training landscapes AI/ML algorithms. We are excited to pursue these opportunities and invite the community to use these methods as a jumping off point for their own interests.

%% file: sections/Acknowledgements.tex
\section{Acknowledgements}
\label{sec:acknolwedgements}

The research described in this paper was conducted under the Laboratory Directed Research and Development Program at Pacific Northwest National Laboratory, a multi-program national laboratory operated by Battelle for the U.S. Department of Energy. Pacific Northwest National Laboratory information release number PNNL-SA-216840.

%% file: appendicies/Wigner_Alternates.tex
\section{Modified Wigner Transforms}
\label{appx:wigner_alternates}

Provided below are several common alternative transformations to the WVT. 

\textit{Gabor-Wigner Transforms:} each use the linear Gabor windowed transformation
\begin{equation}
    G_{\psi}(t,\nu) = \int_{\infty}^{\infty} d \tau \psi(t) e^{-\pi (t - \tau)^2 - i 2 \pi \tau \nu}.
\end{equation}
Gabor-Wigner transforms may refer to any combination of the two namesake transforms~\cite{Pei2007}. The form sampled in this work is
\begin{equation}
    D_{\psi}^{(\alpha,\beta)} (t,\nu) = |G_{\psi}|^{2 \alpha} (t,\nu) \times W_{\psi}^{\beta} (t,\nu),
\end{equation}
where $\alpha$ and $\beta$ are real valued.

\textit{Cohen's Class Transforms:} a convolution of the WVT over the symplectic domain
\begin{equation}
    C_{\psi}(t, \nu) = \left( W_{\psi} * \Pi \right)(t,\nu),
\end{equation}
where $\Pi(t, \nu)$ is the Cohen's kernel function used to smooth the WVT.

\textit{Pseudo-Wigner Transforms:} a type of Cohen's transform of the form
\begin{align}
    &PW_{\psi} (t, \nu) \nonumber \\
    &= \int_{-\infty}^{\infty} d \tau w(\tau/2) w^* (-\tau/2) e^{-2 \pi i \nu \tau} \psi^{*} (t + \tau/2) \psi (t - \tau/2),
\end{align}
where the smoothing kernel only depend on the time domain and is split into a real- or complex-valued window function $w(t)$.

\textit{Polynomial Wigner Transforms:} a generalization of the second order WVT
\begin{align}
    &W^q_{\psi} (t, \nu) \nonumber \\
    &= \int_{-\infty}^{\infty} d \tau e^{-2 \pi i \nu \tau} \prod_{j = 0}^{q/2} \left[\psi (t + c_j \tau/2) \right]^{b_j} \left[ \psi^{*} (t + c_{-j}\tau/2) \right]^{-b_{-j}},
\end{align}
where the order $q$ is an even number, the shift factors $c_j$ are real valued, and the waveform powers $b_j$ are typically integer-valued to maintain the real value of the transform.












%% file: appendicies/Numerical_Methods.tex
\section{Discrete Numerical Methods}
\label{appx:discrete}

Below are details of the numerical methods implemented on the example finite waveforms in Section~\ref{sec:examples}. These waveforms are sampled over time and ordered in a sequence of the form $\mathbf{x} = (x_0, x_1, ..., x_{N-1}) \in \mathbb{R}^N$ is a real-valued vector where each component $x_i$ represents datum associated with an observation time $t_i$, which itself may be organized into a series vector $\mathbf{t} = (t_0, t_1, ..., t_{N-1})$. The waveforms in this work are taken as sampled at equal time intervals, with the numerical methods following suit.

First, the single-dimensional Fourier transform and its inverse may be broken down into the discrete-time Fourier transform (DTFT). These linear and invertable transforms when acting on a generally complex-valued N-sequence take the form
\begin{align}
    F(\mathbf{x})_{\nu_n} &= \sum_{k=0}^{N-1}  x_k e^{- i 2 \pi k n/N}, \\
    F^{-1}(\tilde{\mathbf{x}})_{t_k} &= \frac{1}{2 \pi} \sum_{n=0}^{N-1}  \tilde{x}_n e^{i 2 \pi n k/N},
\end{align}
where the results are also complex-valued N-periodic sequences with elements giving the input series' n-th hamonic mode amplitude for the complex sinusoid wavelets. More commonly used for its computational efficiency and equivalent output is the Fast Fourier Transform (FFT), which is also heavily used in this work where applicable.

The discretized Gabor transform has a similar form to the DTFT, now with a unit amplitude Gaussian window function
\begin{equation}
    G(\mathbf{x})(t)_{\nu_n} = \sum_{k=0}^{N-1}  x_k e^{- (t - k/N)^2/(2 \sigma^2) - i 2 \pi k n/N}.
\end{equation}
The frequency index corresponds to the same harmonic modes as the DTFT, but the time parameter may still lay anywhere on the real line. In this work the time parameter is set to the same interval size as the time series itself, and values set to include the Gaussian's peak. The extent of the Gabor transform in time is the same as the Gaussian window, infinite, however this work chooses a truncation size of the window to $\pm 4 \sigma$ where the Gaussian width is chosen to be an integer multiple of the time interval, thereby limiting the extent in time to $[t_0 - 4 \sigma, t_{N-1} + 4 \sigma]$. Improvements to the computational speed are made using the FFT, series truncation, and an appropriate application of zero-padding of the original time series.

The WVT, pseudo-WVT, and polynomial-WVT Appx.~\ref{appx:wigner_alternates} are similarly discretized over the time-frequency domain
\begin{align}
    W(\mathbf{x})_{t_n, \nu_k} &= \sum_{a=0}^{N-1} e^{- i 4 \pi a k/N} x^{*}_{n + a} x_{n - a}, \\
    PW(\mathbf{x})_{t_n, \nu_k} &= \sum_{a=0}^{N-1} e^{- i 4 \pi a k/N} w_{a} w^*_{-a} x^{*}_{n + a} x_{n - a}, \\
    W^q(\mathbf{x})_{t_n, \nu_k} &= \sum_{a=0}^{N-1} e^{- i 4 \pi a k/N} \prod_{j = 0}^{q/2} \left[x_{n + c_j a}  \right]^{b_j} \left[ x^*_{n + c_{-j} a} \right]^{-b_{-j}},
\end{align}
where the polynomial-WVT displacement parameters $\{c_j\}$ and the power parameters $\{b_j\}$ are typically chosen to be integers. Uses of FFT, series truncation, and zero-padding are also used to improve computational efficiency of the WVT methods.


The remaining computations of entropy and information densities are straightforward applications of point-wise arithmetic operations, discretized integration, and are omitted here.



%% file: appendicies/Autoencoder_Implementation.tex
\section{Autoencoder Implementation Details}
\label{appx:ae}

A standard autoencoder neural network is implemented in this work for the purpose of anomaly detection. The core concept involves learning a lower-dimensional representation that captures the essential features of the training input in an encoder, then attempting to reconstruct the input from the encoded representation using a decoder. The encoder and decoder are the two principle components of the autoencoder network architecture. The network's ability to reconstruct the original input vector measures the quality of the representation.

The encoder's function is to reduce the dimensionality of the input. This is done by using a sequential network consisting of two linear layers, each with a ReLU activation applied after the first linear transformation. 
The output $\mathbf{z}$, set to 8 dimensions, represents the compressed version of $\mathbf{x}$.

The decoder reverses this process, attempting to reconstruct the original input from the compressed representation. It also consists of two linear layers, with a ReLU activation following the first. 
The output $\hat{\mathbf{x}}$ is the reconstructed input.

For training, the Mean Squared Error (MSE) loss function is used, which calculates the average squared difference between the input $\mathbf{x}$ and the reconstructed $\hat{\mathbf{x}}$:
\begin{equation}
{\rm Loss} = {\rm MSE}(\mathbf{x}, \hat{\mathbf{x}}) = \frac{1}{n} \sum_{i=1}^{n} (x_i - \hat{x}_i)^2,
\end{equation}
where $n$ is the dimension of the input vector. For our implementation we use an Adam optimizer and a learning rate of 0.001.

We use \texttt{PyTorch} for our autoencoder implementation. The data is processed in batches of 8, and the model is trained for 10 epochs. The autoencoder is applied using a sliding window approach: training on a baseline, followed by evaluation on subsequent data. This methodology allows for the detection of anomalies by observing increases in the reconstruction error when the evaluation data is not of the same form as the training baseline.
This process enables the identification of anomalies through changes in the reconstruction loss.

%% file: appendicies/Further_Results.tex
\section{Further Results}
\label{appx:further}

\begin{figure}[ht!]
\begin{center}
    \includegraphics[width=1.0\columnwidth]{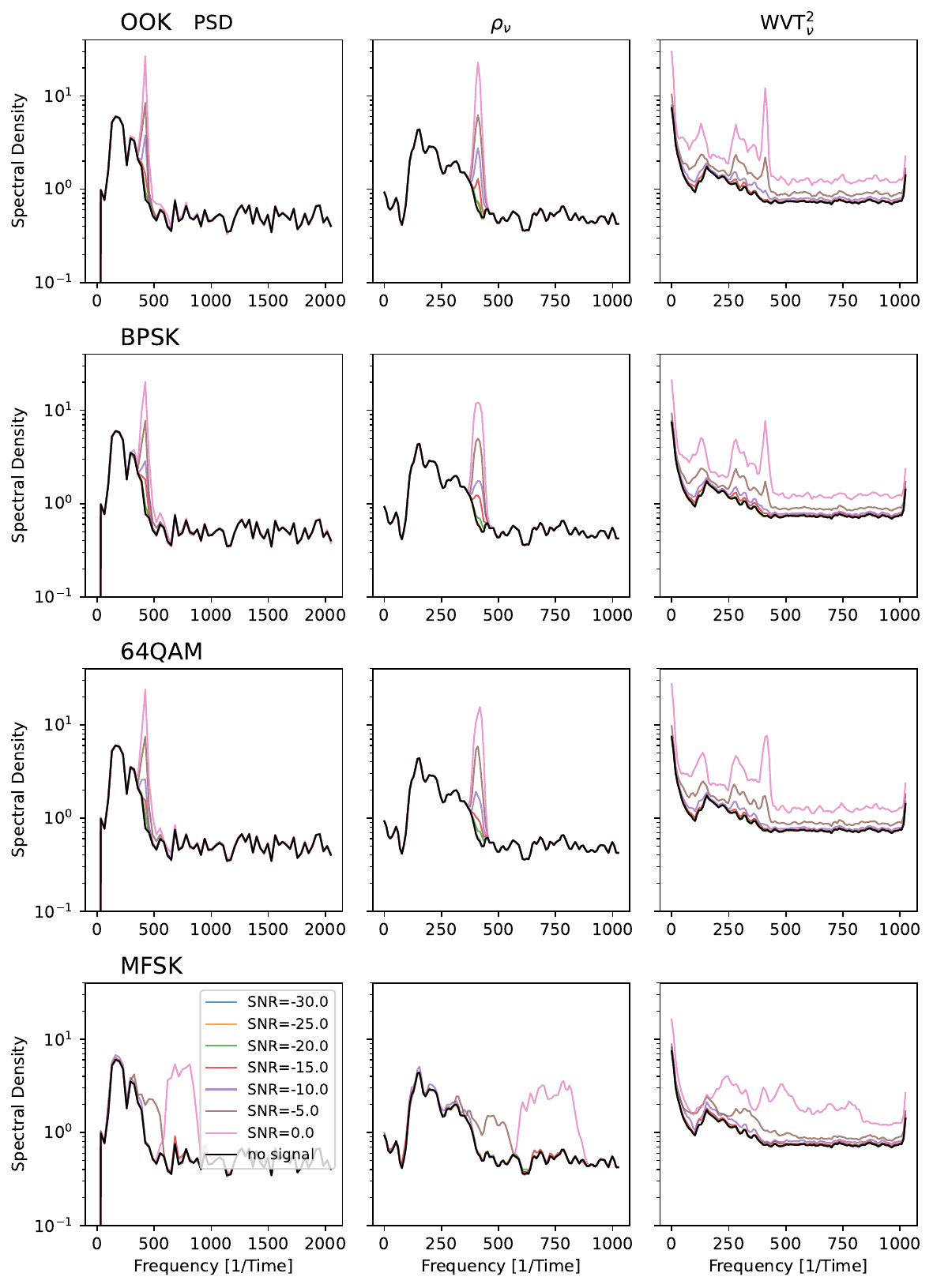}
    \caption{Localized PSD, WVT frequency spectrum $\tilde{\rho}_{\nu}$, and WVT-squared frequency spectrum, computed waveforms with AWGN+MFSK background and a single injection.}
    \label{fig:multi_background_localization}
\end{center}
\end{figure}

\begin{figure}[ht!]
\begin{center}
    \includegraphics[width=1.0\columnwidth]{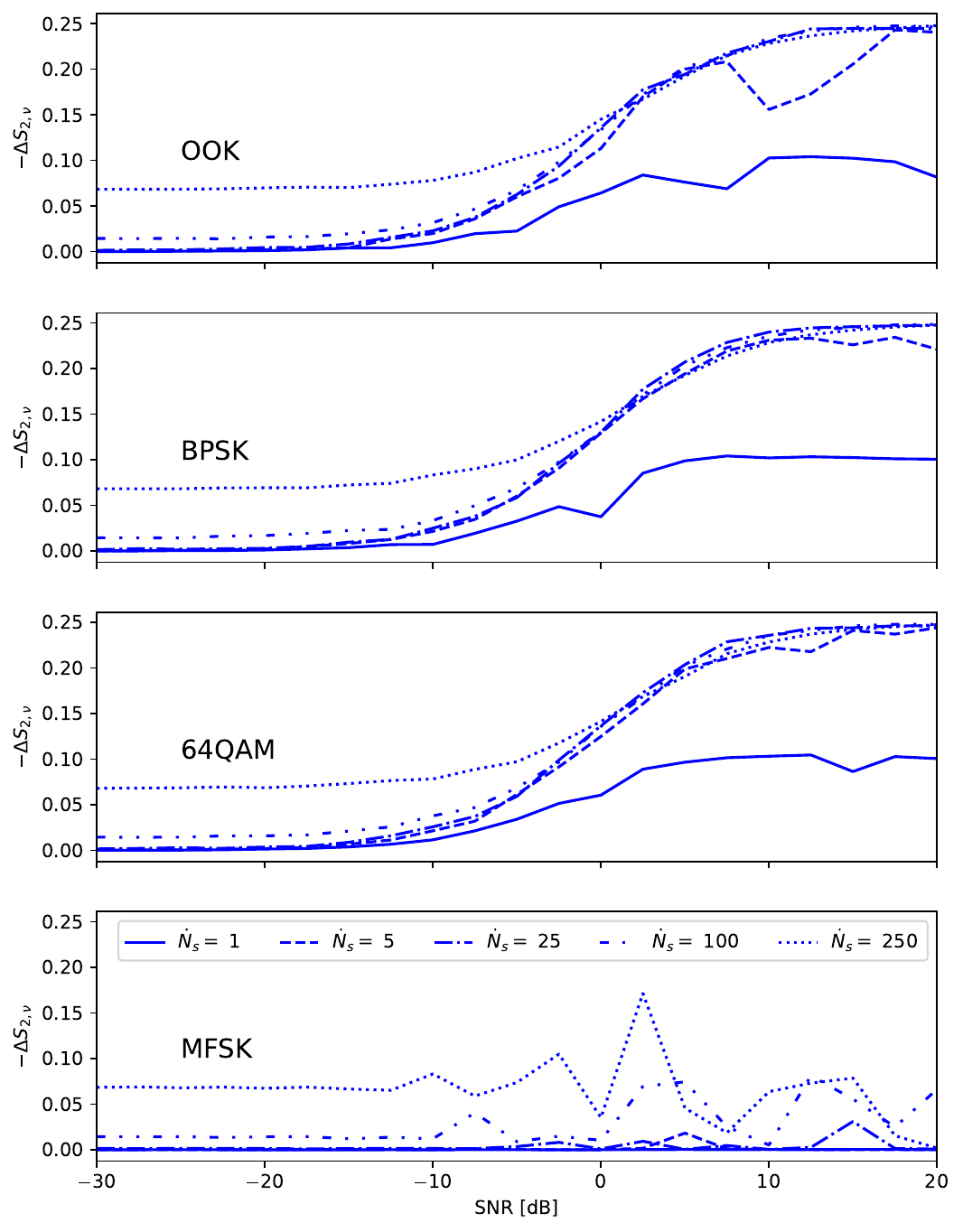}
    \caption{Relative entropy volume for example single-injection signals over AWGN+MFSK background. The relative measurements are made between the carrier mode and a nearby region of the same width. MFSK injections figures are largely unsettled due to lack of a well-defined prescribed center frequency. Note the similar shift of convergence to higher SNR compared to the AWGN background only results.}
    \label{fig:multi_background_volume}
\end{center}
\end{figure}

\begin{figure}[ht!]
\begin{center}
    \includegraphics[width=1.0\columnwidth]{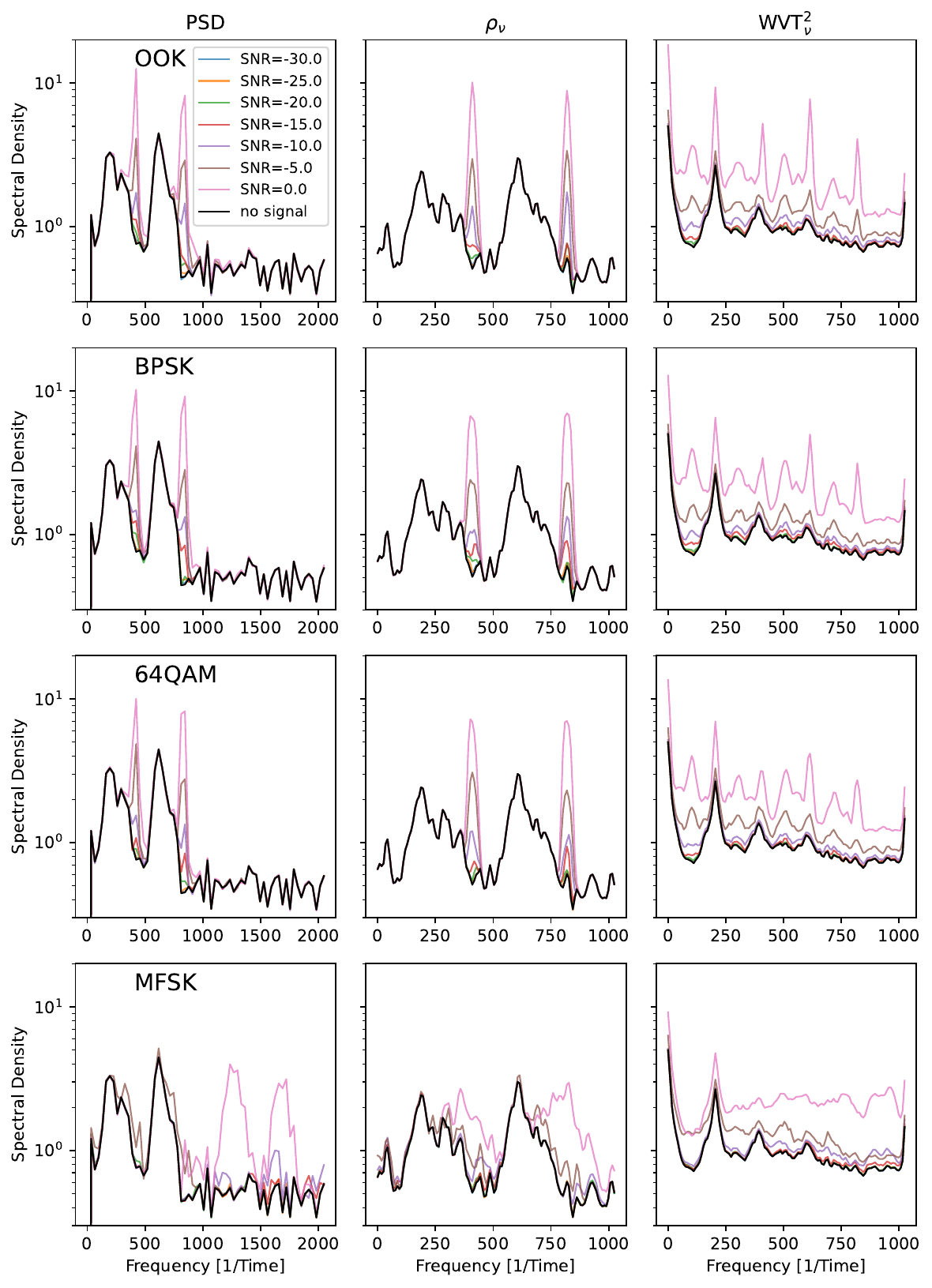}
    \caption{Localized PSD, WVT frequency spectrum $\tilde{\rho}_{\nu}$, and WVT-squared frequency spectrum, computed waveforms with AWGN+(2)MFSK background and two matched injections.}
    \label{fig:double_signal_localization}
\end{center}
\end{figure}

\begin{figure}[ht!]
\begin{center}
    \includegraphics[width=1.0\columnwidth]{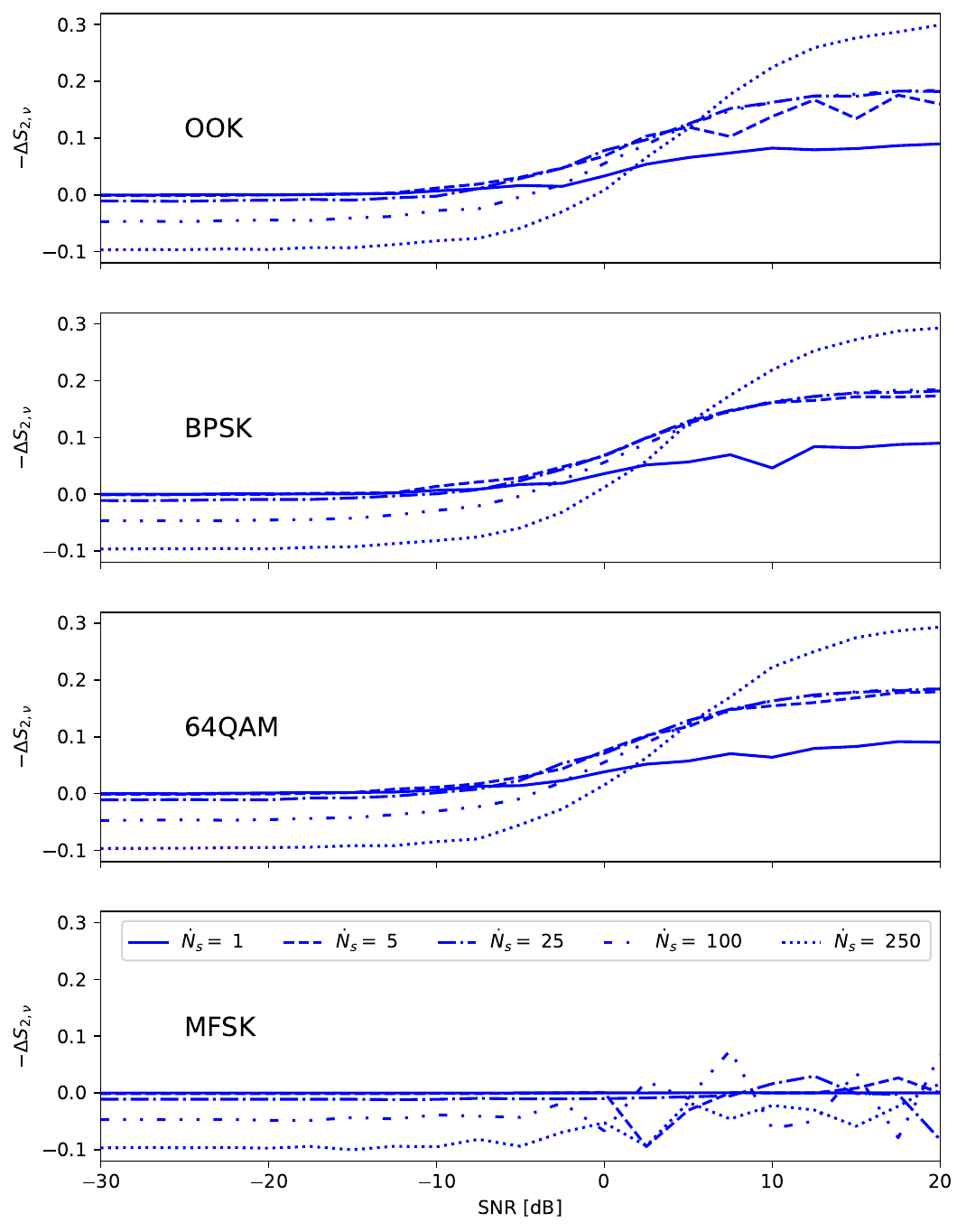}
    \caption{Relative information and entropy volume for example double matched injection signals over AWGN+(2)MFSK background. The relative measurements are made between the carrier mode and a nearby region of the same width. MFSK injections figures are largely unsettled due to lack of a well-defined prescribed center frequency. Note the further degradation in this simple measure due to the increasingly crowded background.}
    \label{fig:double_signal_volume}
\end{center}
\end{figure}

This appendix completes the set of figures discussed though not shown in the main text. The AWGN+MFSK background and single injection demonstration has frequency spectrum examples in Fig.~\ref{fig:multi_background_localization} and data volume estimates in Fig.~\ref{fig:multi_background_volume}. The AWGN+(2)MFSK background and matched injection demonstration has frequency spectrum examples in Fig.~\ref{fig:double_signal_localization} and data volume estimates in Fig.~\ref{fig:double_signal_volume}.


